\documentclass[11pt]{article}
\usepackage{acl}
\usepackage{amsfonts}
\usepackage{amsmath}
\usepackage{amssymb}
\usepackage{amsthm}
\usepackage{bm}
\usepackage{booktabs}
\usepackage{enumitem}
\usepackage[T1]{fontenc}
\usepackage{graphicx}
\usepackage{inconsolata}
\usepackage[utf8]{inputenc}
\usepackage{latexsym}
\usepackage{makecell}
\usepackage{microtype}
\usepackage{multirow}
\usepackage{subcaption}
\usepackage[most]{tcolorbox}
\graphicspath{{figs/}}
\tcbset{
  promptbox/.style={
    enhanced,
    breakable,
    colback=white,
    colframe=black!40,
    boxrule=0.5pt,
    arc=2pt,
    left=6pt,right=6pt,top=6pt,bottom=6pt,
    fonttitle=\bfseries,
    title={#1},
  }
}
\usepackage{times}

\usepackage{listings}
\usepackage{xcolor}
\definecolor{codebg}{HTML}{F8F9FB}
\definecolor{codeframe}{HTML}{4C72B0}
\definecolor{keyword}{HTML}{7B3294}
\definecolor{function}{HTML}{2166AC}
\definecolor{comment}{HTML}{7A7A7A}
\lstdefinestyle{econsim}{
    language=Python,
    basicstyle=\ttfamily\small,
    backgroundcolor=\color{codebg},
    keywordstyle=\color{keyword}\bfseries,
    commentstyle=\color{comment}\itshape,
    stringstyle=\color{orange!70!black},
    showstringspaces=false,
    frame=none,
    columns=fullflexible,
    keepspaces=true,
    xleftmargin=0.5em,
    aboveskip=0pt,
    belowskip=0pt,
}

\newtcolorbox{simbox}[2][]{
    enhanced,
    colback=codebg,
    colframe=codeframe,
    coltitle=white,
    fonttitle=\bfseries,
    title=#2,
    arc=2mm,
    boxrule=0.8pt,
    left=2mm,
    right=2mm,
    top=1.5mm,
    bottom=1.5mm,
    #1
}

\newcommand{\takehiro}[1]{\textcolor{black}{#1}}

\title{EconSimulacra: A Digital Twin Platform of Socio-Economic Systems Powered by LLM Agents}

\author{
 \textbf{Ryuji Hashimoto\textsuperscript{1,2}},
 \textbf{Masahiro Kaneko\textsuperscript{1,3}},
 \textbf{Kentaro Ueda\textsuperscript{1}},
 \textbf{Takehiro Takayanagi\textsuperscript{1,2}},
 \textbf{Kiyoshi Izumi\textsuperscript{1,2}}
 \\
\textsuperscript{1}Simulacra Inc.,
\textsuperscript{2}The University of Tokyo,
\textsuperscript{3}MBZUAI.
\\
\small{
    \textbf{Correspondence:} \href{mailto:email@domain}{r\_hashimoto@simulacra.co.jp}
}
} 

\begin{document}
\maketitle
\begin{abstract}

\takehiro{Real-world social behavior emerges from tightly coupled domains: economic conditions shape mobility and social interactions, while online attention and offline activity feed back into local popularity and consumer behavior. Capturing these feedback loops requires artificial societies in which agents carry experiences from one domain into decisions in another. Large language models (LLMs) provide a promising foundation for such societies. However, existing LLM-based simulators typically model domains in isolation or merely place them side by side.}
To enable such cross-domain interactions, we present EconSimulacra, a multi-agent social simulator that couples consumer economy, mobility, and social networks through a shared internal-state mechanism. In EconSimulacra, experiences accumulated across different domains are stored in memory and transformed into \takehiro{shared internal states (i.e., stress level) connecting heterogeneous domains through individual decision making.}
This design allows agents to reconcile competing demands arising from multiple domains and generate coherent cross-domain behaviors. A case study shows that the shared internal state mechanisms reproduce a nonlinear relationship between online social attention and offline local popularity, illustrating how realistic cross-domain dynamics can emerge within a unified artificial society. The source code\footnote{\tiny{\url{https://github.com/SimulacraBusiness/econsimulacra}}}, package\footnote{\tiny{\url{https://pypi.org/project/econsimulacra/}}}, documentation\footnote{\tiny{\url{https://simulacrabusiness.github.io/econsimulacra/}}}, live demo\footnote{\tiny{\url{https://econsimulacra.onrender.com}}}, and instruction movie~\footnote{\tiny{\url{https://www.youtube.com/watch?v=tGTWAbAN5m8}}} are publicly available.
\end{abstract}

\section{Introduction}
Understanding socioeconomic systems can benefit from considering the interactions among multiple co-evolving social dynamics rather than treating each domain in isolation~\citep{cross_domain_importance1,cross_domain_importance2}. Socioeconomic systems are complex adaptive systems in which diverse domains, such as economic activity, social networks, and human mobility, evolve simultaneously and influence one another~\citep{artificial_society}.

To this end, large language models (LLMs) provide a promising foundation for building multi-domain artificial societies. Conventional agent-based models often rely on manually designed rules tailored to each individual domain, making it difficult to coordinate multiple interacting domains within a single framework~\citep{abm_limits}. In contrast, LLMs can naturally integrate diverse inputs through a shared natural-language representation and generate decisions that holistically account for these heterogeneous factors~\citep{llm_abm_survey1,llm_abm_survey2}. This capability offers a flexible way to model and couple heterogeneous social dynamics within a single artificial society.

\begin{figure*}[tbp]
  \centering
  \includegraphics[width=0.95\linewidth]{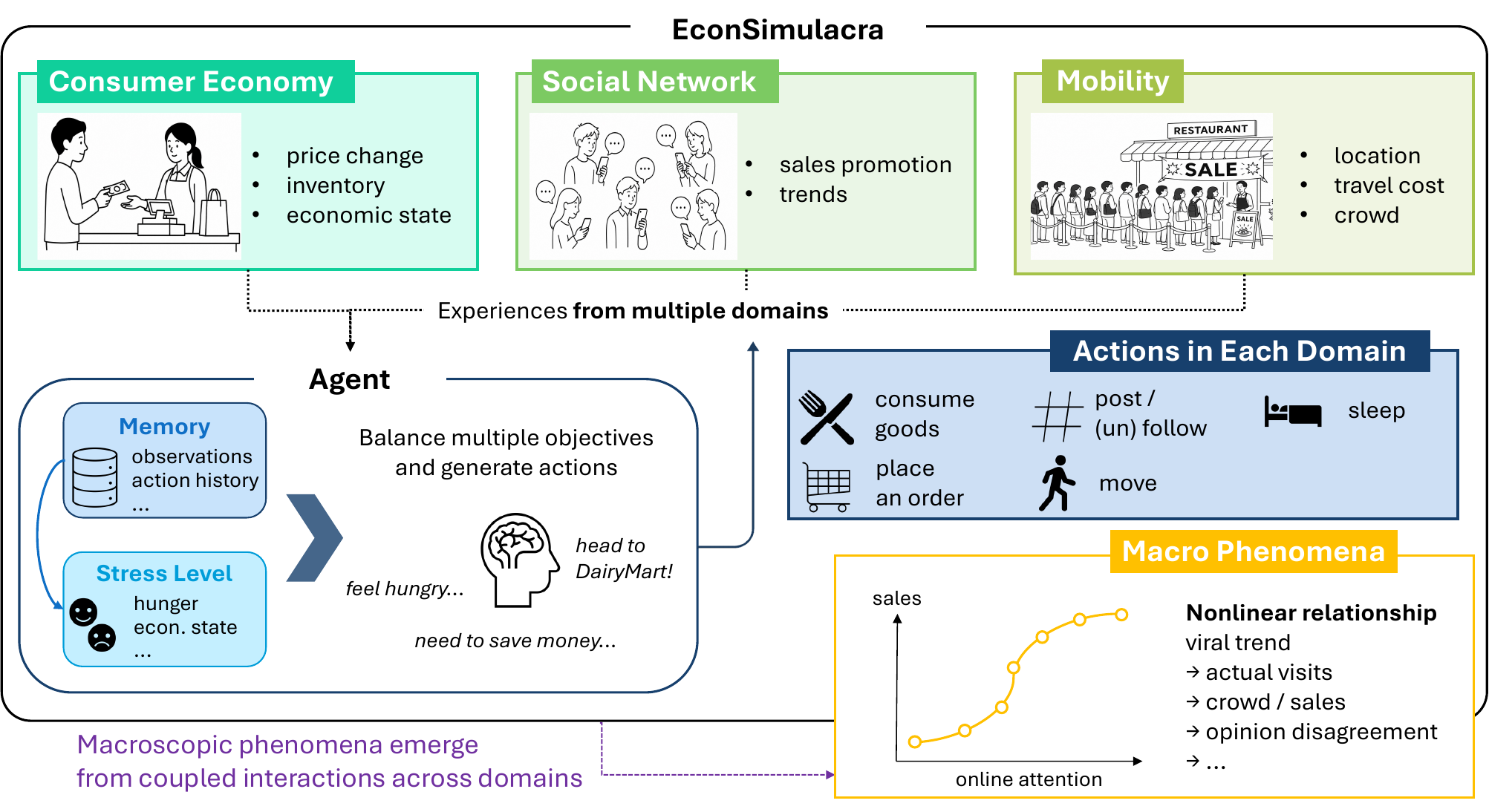}
  \caption{Overview of EconSimulacra. Agents continuously accumulate heterogeneous experiences from the consumer economy, social networks, and mobility domains. These experiences are stored in memory and transformed into shared internal states (i.e., stress level), which guide decisions across domains. The resulting cross-domain interactions generate emergent macroscopic phenomena, such as nonlinear online--offline coupling.}
  \label{Fig:econsimulacra}
\end{figure*}


However, existing LLM-based social simulations either focus on a single domain~\citep{llm_sns1, econagent} or merely place multiple domains within a common environment~\citep{generative_agents,agentsociety}. The mere coexistence of multiple domains does not necessarily imply interaction between them. For cross-domain dynamics to emerge, experiences acquired in one domain must influence behavior in another domain through some shared decision-making process. In real societies, such cross-domain influences are often mediated by internal states arising from physical and social constraints. For example, hunger accumulated through consumption decisions can motivate mobility toward stores, while congestion experienced during travel may subsequently influence social-media activity. Individuals integrate experiences from different domains while balancing physical or social constraints~\citep{multi_finality, constraints_generate_society}. Although prior work~\citep{citysim} has introduced memories and needs, these mechanisms have primarily been used to improve realism within individual domains rather than to generate interactions across domains.

We hypothesize that cross-domain dynamics in LLM societies emerge when heterogeneous experiences are integrated through a shared internal representation. To enable the study of such cross-domain interactions, we propose EconSimulacra, a multi-domain social simulator that couples consumer economy, mobility, and social networks through stress level-based internal states. Agents accumulate experiences across domains in memory, from which stress levels are computed and used as a shared representation of competing demands.

Our experiments demonstrate that introducing stress levels strengthens interactions among heterogeneous social domains and enables the emergence of realistic cross-domain dynamics. In particular, the proposed framework reproduces the nonlinear relationship between online social media activity and offline local popularity, a stylized fact observed in real-world socioeconomic systems~\citep{sns_real_heat}. Ablation experiments further show that removing the stress level-based coupling mechanism weakens these emergent patterns. These findings support the hypothesis that shared internal representations can serve as coupling mechanisms for multi-domain artificial societies.

\section{EconSimulacra Architecture}
\begin{figure}[t]
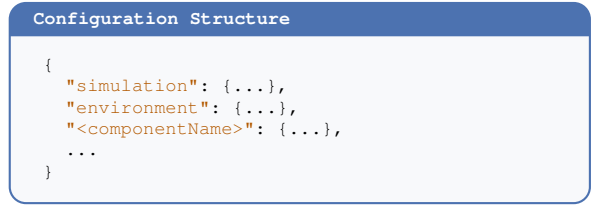

\centering
\begin{simbox}{\scriptsize\texttt{Configuration Structure}}
\begin{lstlisting}[
    style=econsim,
    basicstyle=\ttfamily\scriptsize,
    aboveskip=0pt,
    belowskip=0pt
]
{
  "simulation": {...},
  "environment": {...},
  "<componentName>": {...},
  ...
}
\end{lstlisting}
\end{simbox}
\caption{Structure of the JSON-based configuration file. The configuration specifies the overall simulation settings and environment composition, followed by the definitions of individual components such as agents, spaces, social networks, items, events, and services.}
\label{Fig:config}
\end{figure}

EconSimulacra is an open-source framework for building LLM-based multi-domain social simulations. The framework can be installed via \texttt{pip} and is designed to simplify the development of coupled social simulations. As illustrated in Figure~\ref{Fig:econsimulacra}, the system is organized into two primary layers: \textit{(i) Multi-domain Environments}, which model individual domains such as consumer economy, social networks, and mobility, and \textit{(ii) the Agent Layer}, where LLM-based agents maintain memories of past observations and action histories, convert experiences from multiple domains into a unified stress level representation, and generate actions based on these internal states. 

Simulation scenarios are defined declaratively through a JSON-based configuration file. Figure~\ref{Fig:config} shows the simplified structure of a configuration file. The \texttt{simulation} and \texttt{environment} sections specify the global simulation settings and the composition of the virtual world, while the remaining sections define the individual components (e.g., spaces, social networks, agents, items, events, and services such as LLM API clients and prompt management modules). Since all components are instantiated by name from the configuration, users can seamlessly integrate newly implemented modules alongside the built-in ones without changing the core implementation.

A simulation proceeds by repeatedly collecting observations from the environments, querying each agent for its actions, and updating the environment states based on their actions. Figure~\ref{Fig:simulation_loop} summarizes the overall execution flow. In addition to the Python API, EconSimulacra provides an interactive demonstration interface that enables users to visualize simulation dynamics and inspect the behaviors of individual agents and macroscopic system states.

\begin{figure}[t]
\centering
\begin{simbox}{\scriptsize\texttt{EconSimulacra Simulation Loop}}
\begin{lstlisting}[style=econsim,basicstyle=\ttfamily\scriptsize]
env = Environment(config)
env.reset(seed=42)
num_steps: int
for _ in range(num_steps):
    all_actions = {}
    for agent_id in env.agent_ids:
        agent = env.agent_id2agent[agent_id]
        obs = env.get_observations(agent_id )
        action = agent.act(obs)
        all_actions[agent_id ] = action
    env.step(all_actions)
\end{lstlisting}
\end{simbox}
\caption{Simplified execution flow of a simulation. At each step, agents receive observations from the environments, generate actions, and the environments are updated based on the
joint actions of all agents.}
\label{Fig:simulation_loop}
\end{figure}

\paragraph{\textbf{Multi-domain Environments}}
EconSimulacra models different aspects of social and economic activities as independent environment modules. Each environment maintains its own state transition rules and provides domain-specific observations to agents, while receiving agents' actions to update the underlying world state. This modular design enables multiple domains to interact within a unified simulation while preserving the flexibility to develop and extend each domain independently. Figure~\ref{Fig:econsimulacra} illustrates the three representative environments currently implemented in the framework.
\begin{itemize}[
    leftmargin=1.em,
    itemsep=0.4ex,
    topsep=0.6ex,
    parsep=0pt,
    partopsep=0pt
]
\item \textbf{Consumer Economy:} models purchasing activities and retail markets, providing observations such as product prices, inventories, and economic conditions.
\item \textbf{Social Network:} represents online interactions and information diffusion, including posts, trending topics, sales promotions, and follow/unfollow relationships.
\item \textbf{Mobility:} captures physical movement and location choices, exposing information such as destinations, travel costs, and local crowd levels.
\end{itemize}
At each simulation step, environment modules expose their observations to agents, and subsequently update their internal states according to the actions returned by the agents. Since all environments follow a common interface, new domains can be incorporated by implementing the corresponding observation and transition functions without modifying the existing simulation framework.

\paragraph{\textbf{Agent Layer}}

LLM-based agents serve as the decision-making entities in EconSimulacra by integrating observations from multiple domains and generating actions accordingly. As illustrated in Figure~\ref{Fig:econsimulacra}, each agent maintains two internal states: \textit{Memory} and \textit{Stress Levels}. The \textbf{Memory} module stores a summarized representation of the agent's past experiences across domains, including observations and action histories. The \textbf{Stress Level} module provides a unified internal representation that couples heterogeneous domains. Rather than directly exposing all environment-specific variables to the decision-making process, observations and action histories stored in memory are transformed into a set of interpretable stress levels through rule-based functions. The framework currently provides several built-in stress level models:
\begin{itemize}[
    leftmargin=1.em,
    itemsep=0.4ex,
    topsep=0.6ex,
    parsep=0pt,
    partopsep=0pt
]
\item \textbf{Hunger:} computed from the gap between calorie intake and a target nutritional requirement, where each item can be associated with a calorie value.
\item \textbf{Fatigue:} computed from accumulated travel distance, with stress level alleviation when agents remain at their designated home locations.
\item \textbf{Economic State:} computed from both absolute and relative shortages of available cash and from decreases in owned assets.
\item \textbf{Sleep:} computed from insufficient sleep duration and irregular sleep rhythms.
\item \textbf{Congestion:} computed from the local population density around the agent's current location.
\end{itemize}
Since stress levels are derived from memories through modular rule-based functions, the framework can be easily extended with additional internal drives and domain-specific objectives. The resulting stress level representation enables heterogeneous information from different domains to be handled in a common format, facilitating cross-domain interactions within the agent.

At every simulation step, the LLM receives the summarized memories, current stress levels, and the latest environmental observations, and performs natural-language reasoning to balance multiple objectives, such as satisfying hunger, conserving money, reducing travel costs, or avoiding congestion. The output is a set of executable actions that can be directly applied to the corresponding environments, including consuming goods, placing orders, posting or following users on social networks, moving between locations, and sleeping.

\paragraph{\textbf{Interactive Demo}}

\begin{figure}[tbp]
  \centering
  \includegraphics[width=\linewidth]{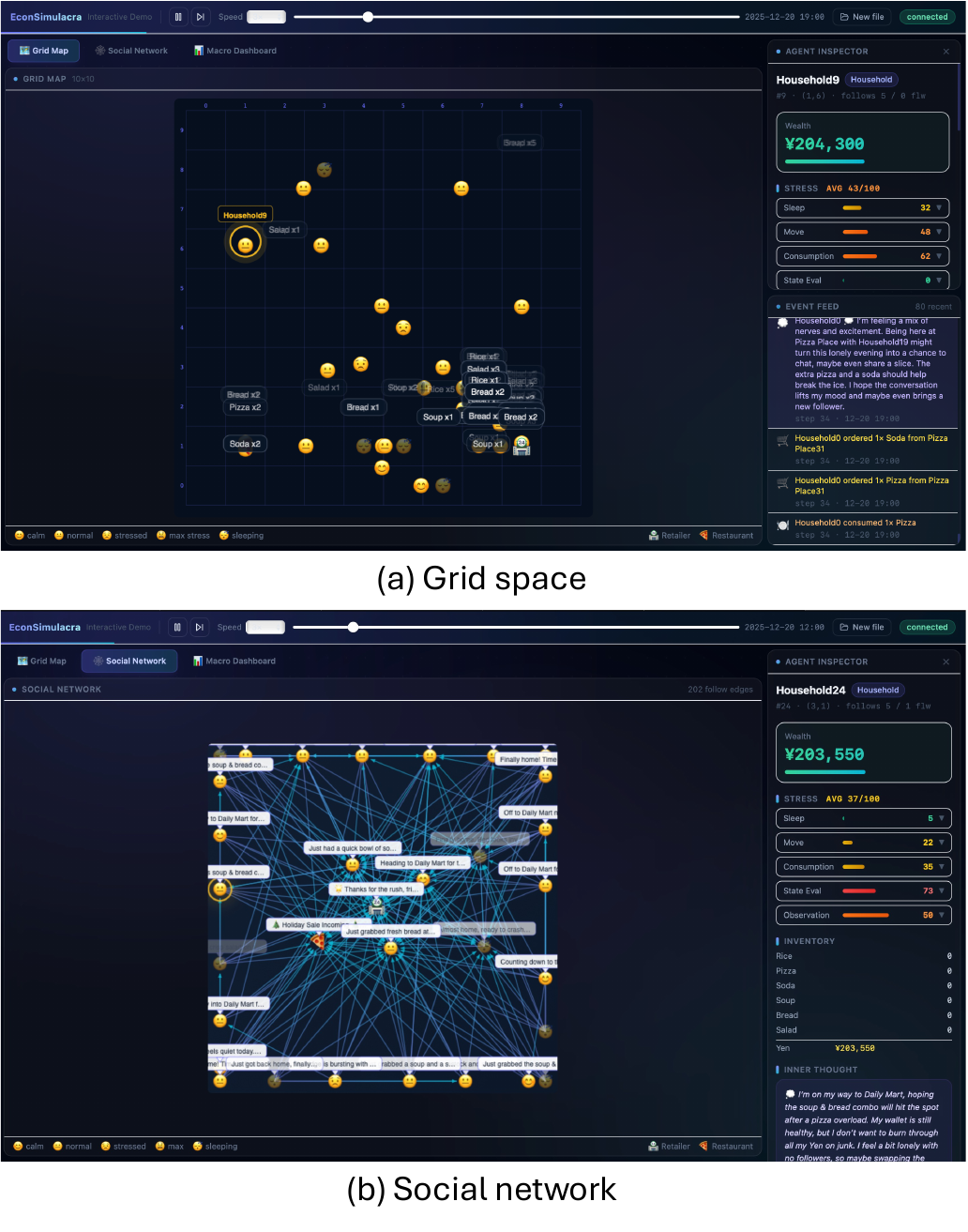}
  \caption{Representative views of the interactive demonstration interface showing (a) agents and physical interactions in grid space, and (b) online communication and information diffusion in the social network.}
  \label{Fig:ui}
\end{figure}

EconSimulacra provides an interactive demonstration interface for visualizing and exploring simulation dynamics. As shown in Figure~\ref{Fig:ui}, the interface allows users to inspect the states of individual domains, such as the grid-based environment, the social network, and macroeconomic indicators, while simultaneously monitoring the internal states and decision-making processes of individual agents. Users can interactively examine agent attributes, memories, stress levels, inventories, and action histories, enabling a detailed understanding of how micro-level decisions emerge from multi-domain experiences. The interface also supports step-by-step playback of simulations, facilitating the analysis and debugging of complex cross-domain interactions.

\section{Experiment}
To demonstrate the ability of EconSimulacra to capture cross-domain interactions, we examine whether coupling among the social network, mobility, and consumer economy domains reproduces the nonlinear relationship between online social attention and offline consumer activity reported in empirical studies~\citep{sns_real_heat}. We test the following hypotheses:
\begin{itemize}[
leftmargin=1.em,
itemsep=0.4ex,
topsep=0.6ex,
parsep=0pt,
partopsep=0pt
]
\item \textbf{H1:} The interaction between online attention and offline consumer activity exhibits a nonlinear relationship.
\item \textbf{H2:} This relationship emerges from \textit{stress level-coupling} across the three domains.
\end{itemize}

To validate the underlying mechanism, we perform an ablation study by comparing two simulation settings: (i) the full EconSimulacra model, where agents accumulate stress levels through experiences across multiple domains and use it for subsequent decision making, and (ii) a stress-disabled variant, where domain-specific experiences do not influence future behaviors through the internal stress levels. By comparing these two conditions, we evaluate whether stress level-coupling is necessary for reproducing the nonlinear coupling between virtual and real spaces.

\paragraph{\textbf{Simulation Setups}} We construct a simplified artificial society consisting of household agents, a rule-based restaurant, and a retailer. The simulation takes place in a grid world, where household agents repeatedly make decisions. As an exogenous intervention, the restaurant launches a predefined discount campaign during a fixed period. Table~\ref{Tab:simulation_setup} summarizes the key simulation parameters. The complete experimental configuration, including simulation settings and execution scripts, is publicly available in the EconSimulacra repository.\footnote{\tiny{\url{https://github.com/SimulacraBusiness/econsimulacra/tree/main/examples/vllm}}} The experiments reported in this paper can be reproduced using the provided configuration files and scripts. The hardware and software configurations used in the experiment are summarized in Appendix~\ref{Sec:computing_infrastructure}.

In the experiment, we conduct 10 independent simulation runs with different random seeds for each of the following LLMs: GPT-OSS-20B~\citep{gpt-oss}, GPT-OSS-120B~\citep{gpt-oss}, Llama 3.1 70B Instruct~\citep{llama3}, and Qwen3.6-35B-A3B~\citep{qwen3}.

\begin{table}[tbp]
\centering
\small
\begin{tabular}{ll}
\toprule
\textbf{Category} & \textbf{Configuration} \\
\midrule
Simulation period & 216 steps (1 hour per step) \\
Grid space & $10 \times 10$ grid world \\
Social network & Follow cap = 5\\
Date time & Dec. 19 09:00 -- Dec. 28 09:00 \\
Household agents & 30 \\
Restaurant & 1 (\textit{Pizza Place}, rule-based) \\
Retailer & 1 (\textit{Daily Mart}, LLM) \\
Restaurant products & Pizza, Soda \\
Retailer products & Rice, Bread, Salad, Soup \\
Event & 40\% pizza discount (Dec. 23--25) \\
Social campaign & Restaurant promotional posts \\
\bottomrule
\end{tabular}
\caption{Simulation setup.}
\label{Tab:simulation_setup}
\end{table}

\paragraph{\textbf{Stress Level-Coupling Ablation}}
To investigate the mechanism underlying the emergence of nonlinear online--offline dynamics, we conduct an ablation study by comparing the following two simulation settings. Both settings share the same simulation environment and differ only in whether past experiences are accumulated as internal stress levels that influence future decision making.
\begin{itemize}[
    leftmargin=1.em,
    itemsep=0.4ex,
    topsep=0.6ex,
    parsep=0pt,
    partopsep=0pt
]
\item \textbf{EconSimulacra (full).} Agents accumulate experiences across multiple domains as internal stress variables, which subsequently influence future decision making. Consequently, experiences in one domain can indirectly affect behaviors in other domains through the shared internal state.
 \item \textbf{EconSimulacra (w/o stress level).} Agents interact with the same environment and perform the same actions as in the full model, but the stress-level abstraction is removed. Instead of maintaining integrated internal stress states, agents directly receive the raw experience histories from each domain. The aggregation and prioritization of these heterogeneous experiences are therefore left entirely to the LLM during decision making.
\end{itemize}
By comparing these two conditions, we evaluate whether stress-mediated memory is necessary for reproducing the nonlinear coupling between online attention and offline consumer activity.

\paragraph{\textbf{Evaluation Metrics}} We quantify the coupling between online attention and offline consumer activity for the restaurant (\textit{Pizza Place}) using the notions of \emph{online heat} and \emph{offline heat}~\citep{sns_real_heat}. In our setting, the offline heat $H_{\mathrm{offline}}\in\mathbb{R}_{\geq0}$, is defined as the daily sales of Pizza Place, while the online heat $H_{\mathrm{online}}\in\mathbb{R}_{\geq0}$, is defined as the daily number of social media posts containing the keyword ``pizza.'' To evaluate whether the simulation reproduces the nonlinear relationship observed in real-world online--offline interactions, we fit the following quadratic regression model:

\begin{align}
H_{\mathrm{offline}}=aH_{\mathrm{online}}^2+bH_{\mathrm{online}}+c.
\end{align}
Here, $a\in\mathbb{R}$ represents the strength of the \emph{saturation effect}, capturing the decrease in offline activity due to factors such as congestion, while $b\in\mathbb{R}$ represents the strength of the \emph{online attention effect}, reflecting the positive impact of social attention on consumer visits. Based on empirical findings, we expect the simulation to exhibit $a<0$ and $b>0$.

In addition, we examine the relationship between the volume and diversity of online opinions by computing the Pearson correlation coefficient between the daily number of ``pizza''-related posts and the standard deviation of their sentiment scores. The sentiment score of each post is estimated using the RoBERTa-based sentiment classifier~\citep{sentiment}. We hypothesize that periods of heightened online attention are accompanied by a greater diversity of consumer experiences and opinions, resulting in a positive correlation between online heat and sentiment dispersion.

\begin{table*}[t]
\centering
\small
\begin{tabular}{lcccc@{\hspace{1.2em}}cccc}
\toprule
& \multicolumn{4}{c}{\textbf{full}}
& \multicolumn{4}{c}{\textbf{w/o stress level}} \\
\cmidrule(lr){2-5}
\cmidrule(lr){6-9}
\textbf{Model}
& $a$ & $b$ & $R^2$ & Corr
& $a$ & $b$ & $R^2$ & Corr \\
\midrule
GPT-OSS-120B           & $-0.337^{**}$ & $1.120^{**}$ & $0.433$ & $0.214^{*}$ & $0.240$  & $0.475$  & $0.400$  & $0.034$ \\
GPT-OSS-20B            & $-0.337^{**}$ & $0.451^{**}$ & $0.349$ & $0.201^{*}$ & $0.003$ & $0.911^{**}$ & $0.712$  & $-0.323$  \\
Llama 3.1 70B Instruct & $-0.416^{**}$ & $0.573^{**}$ & $0.589$ & $0.470^{**}$ & $0.129$ & $0.496^{*}$ & $0.235$  & $-0.010$  \\
Qwen3.6-35B-A3B        & $-0.085^{*}$ & $0.621^{**}$ & $0.374$ & $0.377^{**}$ & $0.042$ & $0.625^{**}$ & $0.428$  & $0.190^{*}$  \\
\bottomrule
\end{tabular}
\caption{Comparison between the EconSimulacra (full) and an ablated variant without stress-level mechanism (w/o stress level). \textit{Note.} * $\text{p}<0.05$, ** $\text{p}<0.01$,}
\label{Tab:stress_ablation}
\end{table*}

\section{Results and Discussion}
Table~\ref{Tab:stress_ablation} provides quantitative evidence supporting \textbf{H1}. Across all tested LLM backbones, the full model consistently exhibits a negative quadratic coefficient ($a<0$), a positive linear coefficient ($b>0$), and a positive correlation between the volume of online discussions and the diversity of sentiment. The comparison between the full model and the ablated variant further supports \textbf{H2}. When the stress-level mechanism is removed, the negative quadratic coefficient becomes substantially weaker, and the correlation between online discussions and sentiment diversity is also greatly reduced. These findings indicate that the nonlinear relationship does not emerge merely because multiple domains coexist, but because information from the social network, mobility, and consumer economy is coupled through the shared internal representation.

A representative cross-domain feedback loop observed in our simulation illustrates the mechanism underlying this result. Promotional information disseminated through the social network attracts customers to a restaurant, changing real-world purchasing and mobility patterns. The resulting experiences, including both positive experiences (e.g., discounted purchases) and negative ones (e.g., congestion), are reflected in agents' stress levels and subsequently influence future online discussions. These updated discussions further amplify or suppress later visits, creating intertwined positive and negative feedback loops spanning the social-network, mobility, and consumer-economy domains. Qualitative examples of these processes are provided in Appendix~\ref{Sec:qualitative_analysis}. Together, these quantitative and qualitative findings suggest that stress-level coupling serves as an effective mechanism for integrating heterogeneous domain-specific constraints, providing a practical design principle for constructing coupled multi-domain social simulations.

\section{Related Work}
\paragraph{\textbf{LLM-based Social Simulations}} Recent studies~\citep{llm_abm_survey1,llm_abm_survey2,llm_abm_survey3,llm_abm_survey4} demonstrate that LLMs provide a foundation for social simulation. Building on these capabilities, LLM-based social simulations have rapidly expanded across a wide range of domains. Existing studies have applied LLM agents to simulate everyday human activities~\citep{generative_agents}, urban and city-scale dynamics~\citep{citysim}, the emergence of social norms~\citep{llm_social_norm}, information diffusion on social media~\citep{llm_sns1, llm_sns2}, opinion formation~\citep{llm_opinions1, llm_opinions2}, macroeconomic behavior~\citep{econagent}, and financial market dynamics~\citep{hashimoto_prima, hirano_prima}. Collectively, these studies highlight the versatility of LLM agents as a common computational substrate for modeling and integrating heterogeneous social and economic domains.

A number of software frameworks have also been developed to support LLM-based social simulations in specific application domains. Representative examples include SmallVille for daily life and social activities~\citep{generative_agents}, CAMEL~\citep{camel}, AgentSims~\citep{agentsims}, and Sotopia~\citep{sotopia} for conversational scenarios, Agent Trading Arena for financial markets~\citep{agent_trading_arena}, and OASIS~\citep{oasis} and MOSAIC~\citep{mosaic} for social media dynamics. More general-purpose platforms such as Concordia~\citep{concordia} and AgentSociety~\citep{agentsociety} provide reusable infrastructures for constructing LLM agent societies across multiple tasks. However, these frameworks primarily focus on modeling a single domain or interaction modality, and largely overlook the explicit coupling of multiple social domains within a unified simulation environment.

\section{Conclusion}
We presented EconSimulacra, an open-source platform for constructing LLM-based artificial societies with coupled consumer economy, mobility, and social-network dynamics. The platform provides a unified framework in which domain-specific observations are integrated into shared internal states and subsequently influence agents' future decisions. Through a case study of online--offline interactions around restaurant promotions, we demonstrated how localized events can propagate across multiple domains and reproduce emergent social phenomena. Our results suggest that reproducing real-world social phenomena requires not only modeling multiple domains, but also providing a mechanism that integrates heterogeneous experiences and constraints into coherent cross-domain feedback loops. We envision EconSimulacra as a step toward a new paradigm in computational social science: moving beyond the isolated analysis of individual domains toward the generative understanding of how interactions between domains give rise to complex social dynamics.

\section*{Limitations}

\paragraph{\textbf{Limited Domain Coverage}}
The current implementation of EconSimulacra focuses on the interactions among three domains: consumer economy, mobility, and social networks. While these domains already capture a variety of online--offline feedback loops, many important components of real-world societies remain outside the scope of the present platform. In particular, labor markets and financial markets represent natural extensions that could introduce additional channels of interaction. Because the platform is designed around modular environments and a shared coupling mechanism, we believe that incorporating such domains can be achieved without fundamentally changing the underlying architecture. Expanding the range of modeled domains is therefore an important direction for building more comprehensive artificial societies.

\paragraph{\textbf{Open Challenges in Cross-Domain Coupling}}
In the current implementation, cross-domain interactions are mediated through a stress level-based internal representation that aggregates heterogeneous experiences and constraints. While our results demonstrate that this mechanism is effective for reproducing coupled online--offline dynamics, stress level should be regarded as only one possible design choice rather than a universal solution. More generally, identifying what kinds of shared internal representations are most suitable for coupling heterogeneous social domains remains an open research question. We therefore view EconSimulacra not only as a simulation platform but also as a testbed for exploring and comparing general mechanisms for cross-domain coupling in artificial societies.

\paragraph{\textbf{Limited Scope of Empirical Study}}
The case study presented in this paper focuses on a specific example of online--offline coupling around restaurant promotions, and therefore represents only a small subset of the social phenomena that the platform may support. Establishing the generality of the proposed framework will require applying it to a broader range of domains and validating the resulting dynamics against diverse empirical observations. More fundamentally, we do not view EconSimulacra as a complete model of society, but rather as an experimental infrastructure for studying how interactions among heterogeneous domains generate complex social phenomena. We hope that such LLM-based artificial societies will complement conventional data-driven analyses by enabling researchers to formulate, simulate, and test generative hypotheses about cross-domain social dynamics.

\bibliography{custom}

\begin{figure}[tbp]
  \centering
  \includegraphics[width=\linewidth]{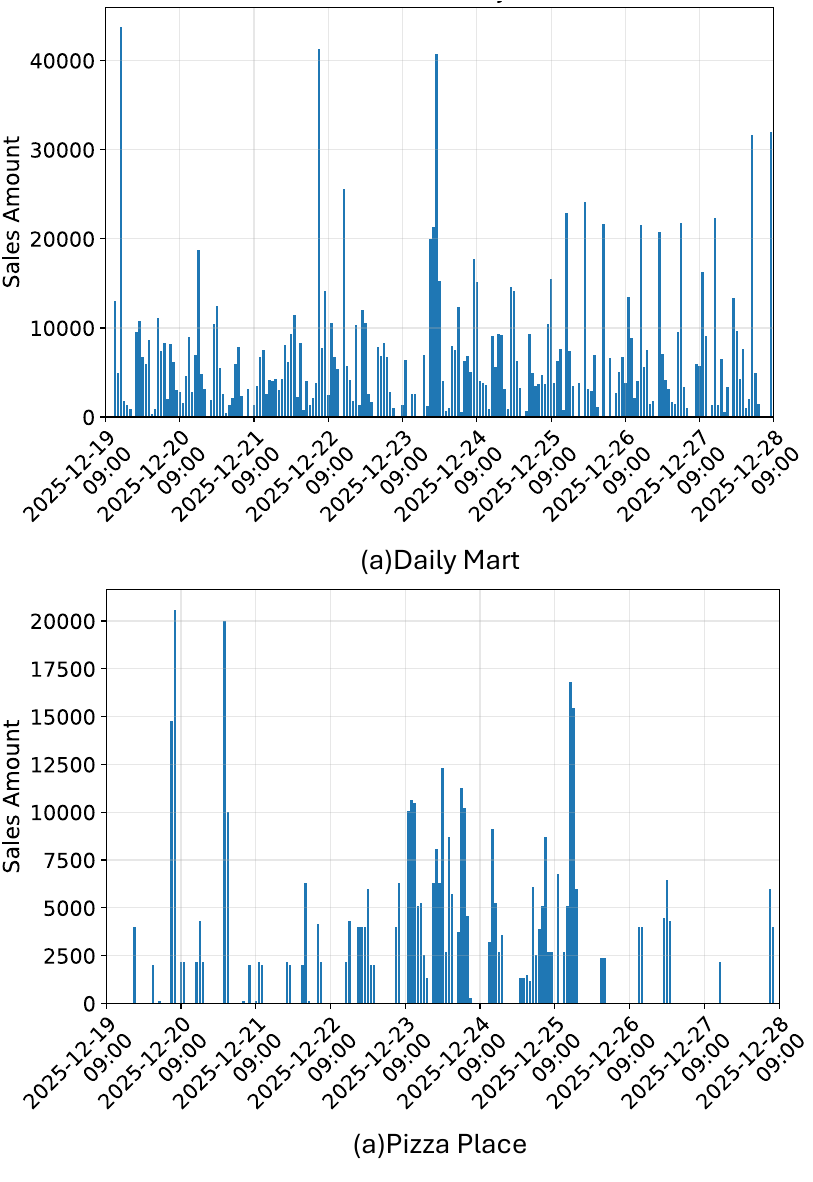}
  \caption{Time series of sales amounts for Pizza Place and Daily Mart in a representative simulation run with GPT-OSS-120B. A promotional campaign was conducted at Pizza Place during December 23--25.}
  \label{Fig:sales}
\end{figure}

\begin{figure*}[tbp]
    \centering

    \begin{subfigure}[t]{0.44\textwidth}
        \centering
        \includegraphics[width=\linewidth]{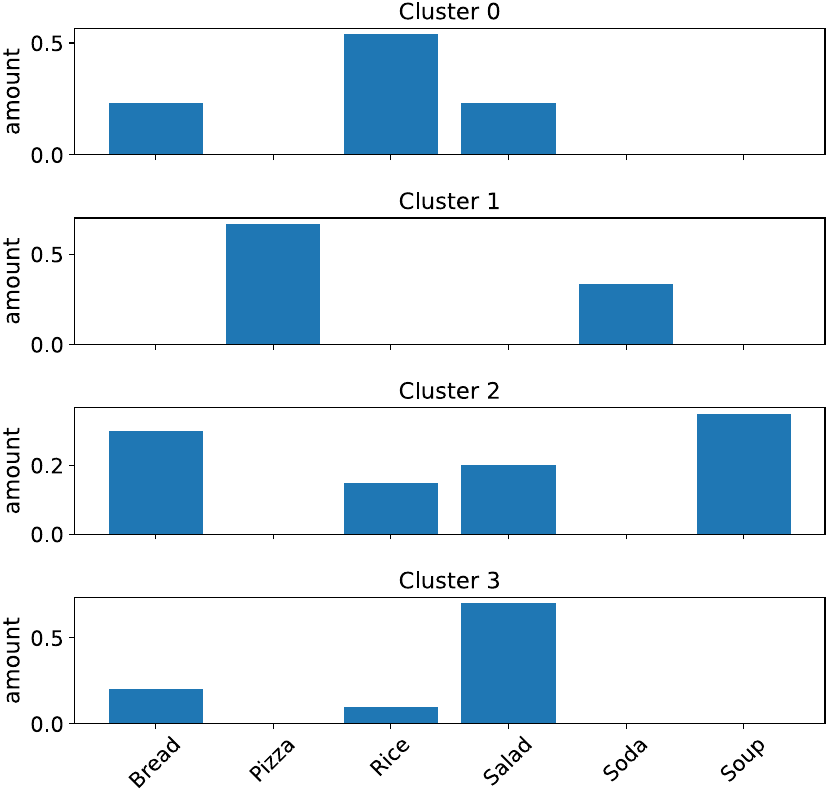}
        \caption{Representative consumption vectors for the clusters.}
        \label{Fig:cluster_representatives}
    \end{subfigure}
    \hfill
    \begin{subfigure}[t]{0.53\textwidth}
        \centering
        \includegraphics[width=\linewidth]{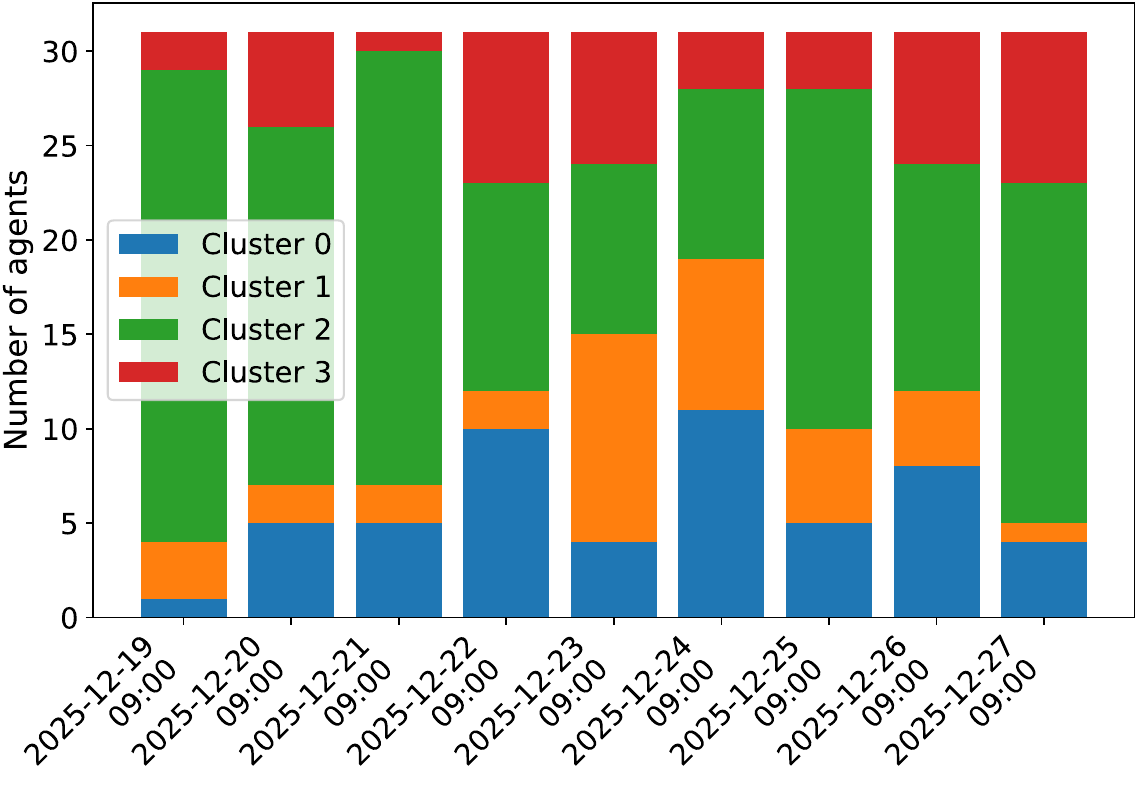}
        \caption{Time series of the number of agents assigned to each cluster.}
        \label{Fig:cluster_counts}
    \end{subfigure}

    \caption{Behavioral clusters derived from normalized daytime consumption vectors and their temporal evolution in the representative GPT-OSS-120B simulation run.}
    \label{Fig:behavior_clusters}
\end{figure*}

\appendix

\section{Computing Infrastructure}\label{Sec:computing_infrastructure}

The experiment is conducted on a workstation equipped with the following hardware and software configurations:

\begin{itemize}
\item \textbf{CPU}: AMD Ryzen 9 3950X 16-Core Processor (32 threads, base clock 3.5 GHz, max boost 4.76 GHz)
\item \textbf{GPU}: 2 × NVIDIA RTX 6000 Ada Generation (48 GB VRAM each)
\item \textbf{Operating System}: Ubuntu 22.04.5 LTS (Jammy Jellyfish)
\end{itemize}

\section{Qualitative Analysis of an Emergent Online--Offline Coupling}\label{Sec:qualitative_analysis}
To complement the quantitative analysis in the main text, we provide a qualitative case study of a single representative simulation run using GPT-OSS-120B. We focus on the three-day promotional event held at Pizza Place during December 23–25 and examine how changes in agent behaviors and online discussions jointly give rise to the observed online–offline coupling.


\paragraph{\textbf{Promotional Event and Aggregate Sales Dynamics}}
Figure~\ref{Fig:sales} shows the resulting sales amount trajectories of the two stores. During the promotional period (December 23--25), Pizza Place exhibits a pronounced increase in sales activity, whereas Daily Mart remains within its ordinary range of fluctuations. This contrast confirms that the promotional campaign acts as a localized exogenous perturbation to the consumer economy domain. In the following analysis, we trace how this local offline event is reflected in the collective behaviors and online discussions of household agents, providing a qualitative illustration of the feedback loop across consumer activities, mobility, and social interactions that underlies the nonlinear online--offline relationship reported in the main text.

\paragraph{\textbf{Behavioral Adaptation of Household Agents}}
To characterize the collective adaptation of household behaviors, we cluster agents according to their normalized daytime consumption-item vectors.\footnote{Specifically, for each agent and each day, we construct a consumption vector by counting purchases of each item category during daytime activities. For example, a daily consumption record of (Bread: 5, Pizza: 2, Rice: 0, Salad: 5, Soda: 1, Soup: 0) is represented as the six-dimensional vector $(5,2,0,5,1,0)$, which is then normalized to unit lengh before clustering. We conduct KMeans clustering with $K=2,4,6,8,$ and $10$, and report the result with the highest adjusted Rand index (ARI) of $0.681$.} Figure~\ref{Fig:behavior_clusters} shows the resulting representative consumption patterns and the temporal evolution of the number of agents assigned to each cluster. The identified clusters capture distinct consumption tendencies, including a pizza- and soda-oriented dining pattern (cluster~1), rice- and bread-centered routine shopping patterns (cluster 0 and 2), and a salad-oriented light-meal pattern (cluster~3). During the promotional period, the population of the pizza-oriented cluster (cluster~1) increases markedly, while the relative sizes of the routine shopping clusters decrease. These observations suggest that the promotional event reshapes the distribution of consumption behaviors across the population.

\begin{figure}[t]
    \centering
    \includegraphics[width=0.98\linewidth]{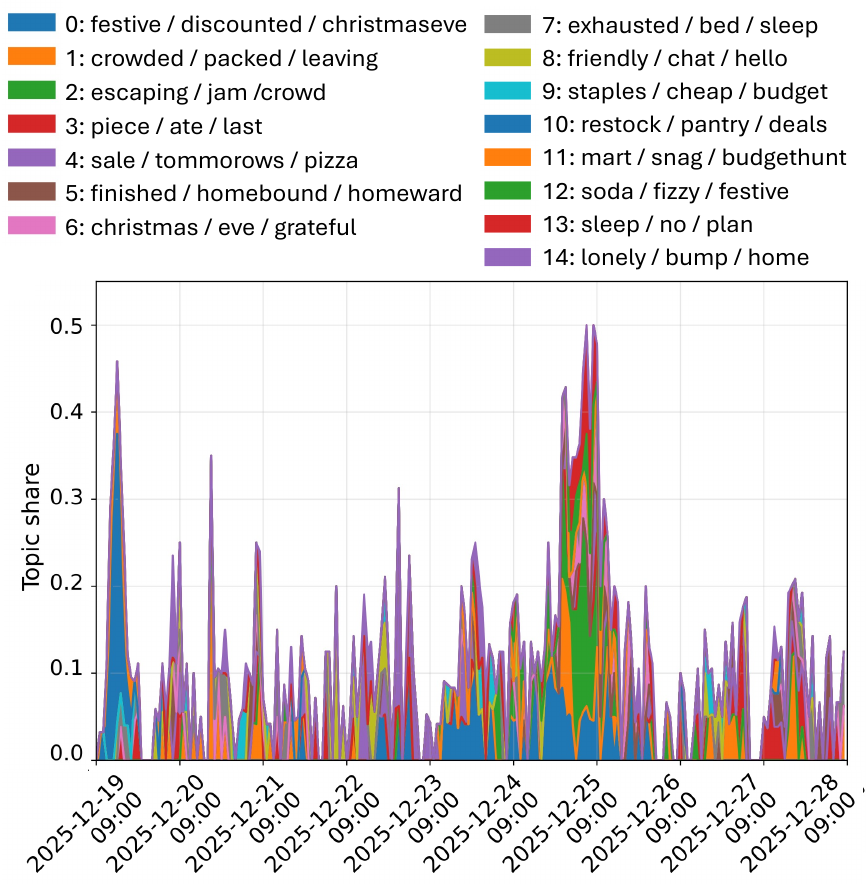}
    \caption{Time series of the topic composition of online discussions in the representative GPT-OSS-120B simulation run. Topics are obtained by applying BERTopic to all posts generated during the simulation, and the figure shows the proportion of each topic over time.}
    \label{Fig:topic_share}
\end{figure}

\paragraph{\textbf{Emergence of Online Discussions}}
To investigate how changes in offline activities are reflected in the social network domain, we extract discussion topics from the corpus of all generated posts using BERTopic~\citep{bertopic}. Figure~\ref{Fig:topic_share} shows the temporal evolution of the resulting topic shares throughout the simulation. During the promotional period, topics directly related to the campaign, such as \textit{``festive / discounted / christmaseve''} (topic~0), become prominent, indicating that information about the sale is actively propagated through online interactions. At the same time, topics associated with the consequences of increased offline activity, including \textit{``crowded / packed / leaving''} (topic~1) and \textit{``escaping / jam / crowd''} (topic~2), also emerge and spread among agents. These observations suggest that online discussions reflect not only the promotional event itself but also the collective experiences generated by the resulting changes in purchasing and mobility behaviors. In this way, the social network domain contains both positive feedback, where information about the sale attracts additional consumers, and negative feedback, where reports of congestion discourage further visits. This coexistence of reinforcing and balancing feedback qualitatively illustrates the cross-domain interaction mechanism underlying the nonlinear online--offline relationship reported in the main text.

\end{document}